\documentclass[aps,amsmath,floats,twocolumn,superscriptaddress]{revtex4}
\usepackage{graphicx}
\usepackage{dcolumn}
\usepackage{multirow}

\newcommand{\species}[1]{\textit{#1}}

\newcommand{\file}[1]{\texttt{#1}}
\newcommand{\website}[1]{\texttt{#1}}

\newcommand{\transreg}{transcriptional regulation}
\newcommand{\trn}{\transreg\ network}

\newcommand{\etal}{\emph{et al}}
\newcommand{\Ecoli}{\species{E. coli}}

\newcommand{\Lrand}{L_{\mathrm{rand}}}
\newcommand{\LDNA}{L_{\mathrm{DNA}}}

\newcommand{\bp}{\mathrm{bp}}
\newcommand{\kbp}{\mathrm{k}\bp}
\newcommand{\Mbp}{\mathrm{M}\bp}

\newcommand{\mysection}[1]{%
  \vspace{12pt}%
  \noindent{\bf\sf#1}\\
  \noindent%
}

\begin{document}

\title{Statistical analysis of the spatial distribution of operons in
the \trn\ of \species{Escherichia coli}}

\author{P. B. Warren}
\affiliation{FOM Institute for Atomic and Molecular Physics, 
Kruislaan 407, 1098 SJ Amsterdam, The Netherlands}
\affiliation{Unilever R\&D Port Sunlight, Bebington, Wirral, CH63 3JW, UK}

\author{P. R. ten Wolde} 
\affiliation{FOM Institute for Atomic and Molecular Physics, 
Kruislaan 407, 1098 SJ Amsterdam, The Netherlands}
\affiliation{Division of Physics and Astronomy, Vrije Universiteit, 
De Boelelaan 1081, 1081 HV Amsterdam, The Netherlands}

\date{October 16, 2003 \file{art-oct-v5.tex}}



\maketitle

\mysection{
We have performed a statistical analysis of the spatial distribution
of operons in the \trn\ of \species{Escherichia coli}. The analysis
reveals that operons that regulate each other and operons that are
coregulated tend to lie next to each other on the genome. Moreover,
these pairs of operons tend to be transcribed in diverging
directions. This spatial arrangement of operons allows the upstream
regulatory regions to interfere with each other. This affords
additional regulatory control, as illustrated by a mean-field analysis
of a feed-forward loop. Our results suggest that regulatory control can
provide a selection pressure that drives operons together in the
course of evolution.
}

Most, if not all, organisms can respond and adapt to a changing
environment. To this end, they can detect, transmit, and amplify
environmental signals, as well as integrate different signals to
perform computations analogous to electronic devices. Indeed, all
organisms can be considered to be information processing
machines. Yet, how the living cell accurately processes information,
is still poorly understood. Recent technological developments,
however, have made it possible to acquire information on the
regulatory architecture of the cell on a massive scale, and extensive
databases are now available that catalog biochemical networks. This
offers unprecedented possibilities to unravel the design principles by
which organisms process information.

The current richness of genomic data surrounding \species{Escherichia
coli} makes it no doubt one of the best characterized of all living
organisms.  The condensation of genes into operons and the
organization of operons into the \trn\ are now well mapped, and this
information has been used to investigate generic features such as the
appearance of motifs in the \trn\ \cite{SOMMA}.  Here, we present a
study of the {\em spatial} organization of operons in the \trn\ of
\Ecoli. Our analysis of the spatial distribution of {\em operons}
provides two distinct advantages over previous studies on the spatial
distribution of {\em
genes}~\cite{Tamames97,Huynen98,Overbeek99,Snel00,SMSV,DeDaruvar02,Ouzounis03}:
firstly, it excludes correlations from genes that belong to the same
operon. Secondly, and more importantly, by focusing on the
higher-level organisation of operons into the \trn, the analysis allows
us to elucidate spatial correlations associated with regulatory
control, for instance, by identifying coregulated pairs of operons
that are adjacent on the DNA.

We find that there is a marked tendency for operons that are related
to each other in the \trn\ to be nearest neighbours, compared to
networks in which operons are randomly assigned positions on the DNA.
Furthermore, the separations between neighbour pairs have a strong
bias towards short distances, which is most pronounced for pairs that
are transcribed in diverging directions. In fact, our analysis
identifies a new, spatial network motif that consists of pairs of
{\em overlapping operons} - operons of which the upstream regulatory
domains overlap.

Several mechanisms could give rise to the observed distributions. The
strong bias towards short separations could be a result of the
mechanisms by which genes and connections between genes arise and
disappear during evolution.  In contrast, it is also conceivable that
there is a functional benefit for having certain operons close to each
other. This would lead to a selection pressure for shorter separations
between certain operons. We do not investigate these scenarios in
detail here, but our data does support the latter scenario for the
diverging neighbour pairs. In particular, we examine a network motif
that has been identified by our statistical analysis: a feed-forward
loop in which the `downstream' operons are transcribed in diverging
directions. The analysis shows that overlapping regulatory domains for
these operons can strongly enhance the response of the network.
Hence, our results suggest that regulatory control can provide an
evolutionary driving force for the formation of overlapping operons.

\mysection{Methods}
Our starting point is the \trn\ data compiled by Shen-Orr \etal\ for
their motif analysis \cite{SOMMA}.  We have annotated their list of
operons with the start- and end-points of the coding regions,
extracted from the various databases of genomic information for
\Ecoli\ \cite{note1,ecocyc}.  We work with coding regions rather than
promoters because the former are easy to identify and the distance
between a promoter and the start of the coding region is small
compared to the typical distances we consider here.  The \trn\
contains 404 operons with 558 links.

We focus on the statistics of the pair separations between operons
which are related in the \trn\ by various definitions (see below), and
on pairs of neighboring operons on the genome. We define the pair
separation to be the distance along the DNA between the coding regions
(an alternative definition as the distance between the starting points
for transcription was explored with similar results).  One basic tool
is the cumulative distribution function $F(s) = \int_0^s ds^\prime
P(s^\prime) $, where $P(s) ds$ is the probability that the distance
between two operons along the DNA has a value between $s$ and $s+ds$.
The cumulative distribution function $F(s)$ is used in preference to
the probability distribution function $P(s)$ because it is readily
visualized even for sparse data sets, and does not need to be
corrected if a logarithmic axis is used for $s$.

\begin{figure}
\includegraphics[width=8.5cm]{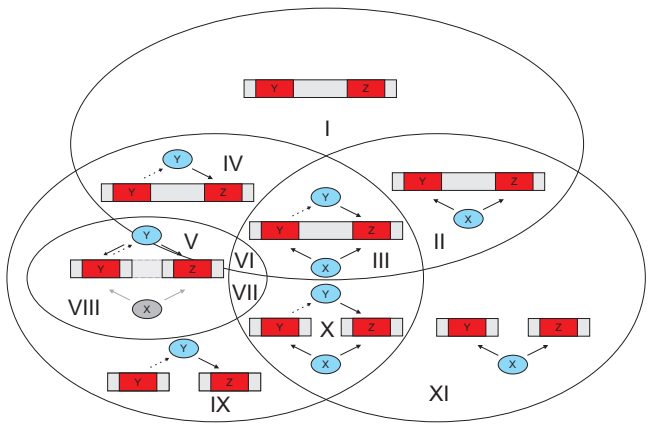}
\caption[]{Schematic overview of the sets of network motifs studied in
this paper.  The full set is given by the ensemble of all possible
operon pairs.  We have split the full set into the following subsets:
the ensemble of nearest neighbour pairs (NN), corresponding to regions
I-VI; the set of pairs that are connected in the \trn\ (TRN),
corresponding to regions III-X; the set of pairs that are
coregulated by a distinct third operon (CR), corresponding to
II+III+VI+VII+X+XI; the set of autoregulated pairs (AR), which is a
proper subset of the TRN set (regions V-VIII; here, protein X only
regulates operon Y and Z in regions VI and VII).  The union of TRN and
CR sets (subsets III+VI+VII+X) corresponds to feed-forward loops
(FFLs), a network motif identified by Shen-Orr
\etal~\cite{SOMMA}.\label{fig:overview}}
\end{figure}

We base our analysis on the operon pairs in three overlapping sets:
pairs of operons that are nearest neighbours (NN) on the DNA, pairs of
operons in the \trn\ (TRN), and pairs of operons that are coregulated
(CR) by a third operon.  In addition, we have also considered
autoregulated (AR) pairs which are TRN pairs for which the controlling
operon also regulates itself.  The different sets and subsets are
schematically indicated in Fig.~\ref{fig:overview}, and the sizes of
the sets in Table~\ref{tab:overview}.

In order to determine the statistical significance of the different
quantities for the \Ecoli\ network, we have calculated the
corresponding expectation values for an ensemble of random networks.
Since we are primarily interested in the network motifs that arise due
to the spatial organisation of the network and not in those that are a
consequence of the topology of the network (which have already been
identified by Shen-Orr \etal~\cite{SOMMA}), we define a random network
to be a network with a connectivity of that of the \Ecoli\ network,
but with a random assignment of operon positions and orientations on
the genome. Hence, not only the NN set of the \Ecoli\
network and a random network are equal in size, also the sizes of the
TRN, CR and AR sets in the \Ecoli\ network equal those in a random
network, as the sizes of these sets are determined by the topology of
the network. In contrast, the sizes of the unions corresponding to
regions II-VI (see Fig.~\ref{fig:overview}) differ between the \Ecoli\
network and a random network, since they are determined by the spatial
distribution of the operons.  For the \Ecoli\ network, we can directly
obtain the sizes of the respective unions from the various
databases~\cite{note1,ecocyc}. For the ensemble of random networks, we
have calculated the expectation value for the number of pairs,
$M_\alpha$, in the union formed by the overlap of the set of nearest
neighbour pairs with set $\alpha$ (be it set TRN, AR or CR), using
$M_\alpha = p N_\alpha$, where $p$ is the probability that a randomly
chosen pair is, in fact, a nearest neighbour pair and $N_\alpha$ is
the number of pairs in set $\alpha$. We have verified these
calculations by generating random networks and computing the
quantities directly.  The $P$-values reported are probabilities of
finding a subset in the random network of at least the size as
observed in \Ecoli, computed using the same statistical model.

\begin{figure}[t]
\includegraphics[width=8.5cm]{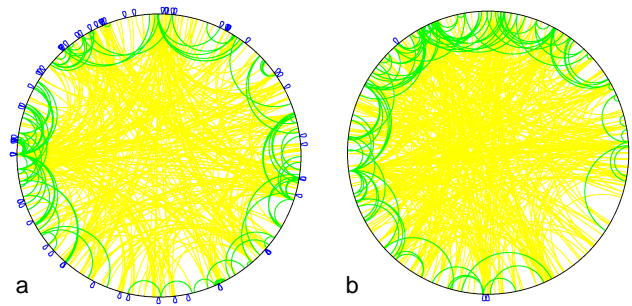}
\caption[]{The \trn\ of \Ecoli\ shown as links between operons on the
genome.  Maps are shown for (a) the real \Ecoli\ network and (b) a
representative `randomised' network with the same topology, but with a
random permutation of the positions of the operons. The color code
is: blue: $s < 10\,\kbp$; green: $10\,\kbp < s < 500\,\kbp$; yellow:
$s > 500\,\kbp$. Note the much greater prevalence of the `short'
distances in the real map (a) compared to the randomised map
(b).\label{figm}}
\end{figure}

\begin{figure}
\includegraphics[width=8.5cm]{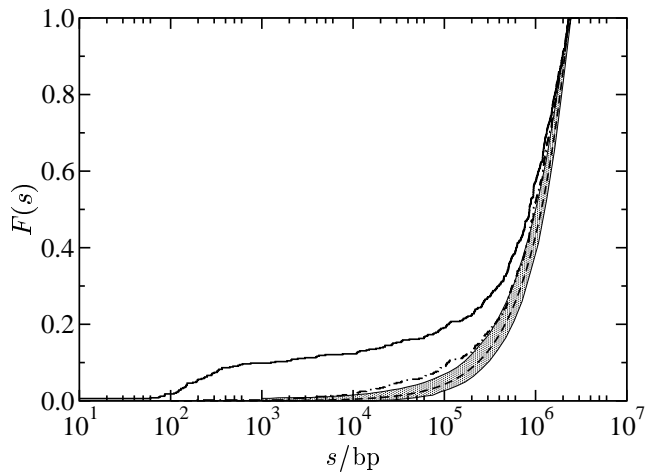}
\caption[]{The cumulative probability distribution $F(s)$ for the
distances along the DNA between operons in the transcriptional
regulatory network (TRN) of \species{E. coli} (solid line). The dashed
line corresponds to the average of an ensemble of networks that have
been obtained by randomly permuting the positions of the operons, thus
preserving the network topology; the grey area denotes the 98\%
confidence regime. The dashed line is indistinguishable from the
distribution function that is expected for a random network, which is given by
$F(s) = s/(\LDNA/2)$. Note the significant 10--15\% fraction with
$s\alt 1\,\kbp$ for the \Ecoli\ network.  The dotted-dashed line
corresponds to the TRN pairs that are \emph{not} nearest neighbours.
The mean lengths of genes and operons in the data set are $1.0\,\kbp$
and $2.1\,\kbp$, respectively.\label{fig:Ftrn}}
\end{figure}

\mysection{Results}
\mysection{Transcriptional Regulation Network}
In fig.~\ref{figm} we compare a `map' of the real transcriptional
regulation network to a map of a randomised version of the network;
the random network has been obtained by randomly permuting the positions of
the operons on the DNA, thus preserving the topology of the network. It
is seen that the real network exhibits a larger number of `short'
(blue) links and a smaller number of `long' (yellow) links, as
compared to the random network. This indicates that operons that
regulate each other tend to lie closer to each other than can be
expected for a random network. We can quantify this by calculating the
distribution functions for the distances between the network
pairs. Fig.~\ref{fig:Ftrn} shows the cumulative disitribution
functions for the \Ecoli\ network and for the ensemble of random
networks.  For the ensemble of random networks, we expect that the
separation between network pairs is uniformly distributed between zero
and $\LDNA/2 = 2.3\,\Mbp$, where $\LDNA = 4.64\,\Mbp$ is the total
length of the DNA. Fig.~\ref{fig:Ftrn} shows that this is indeed the
case. For the real network, however, we find marked deviations from a
uniform distribution. It is seen that up to $\approx 200\,\bp$, the
cumulative distribution function is close to zero; this lower cut-off
reflects the typical sizes of promoter regions for operons.  However,
after $\approx 200\,\bp$ the cumulative distribution function of the
\Ecoli\ network sharply increases by some 10--15\%, until it follows a
nearly uniform distribution from $\approx 1\,\kbp$ upwards.

What is the origin of the pronounced increase in $F(s)$ at around
$200\,\bp$? Do operons that are linked in the \trn\ tend to be nearest
neighbors? Table~\ref{tab:overview} shows that is indeed the case. In
\Ecoli, 55 out of the 497 \trn\ pairs are nearest neighbours, as
compared to the average of $2.5$ in the ensemble of random
networks. Moreover, these (NN,TRN) pairs tend to lie very close to
each other: out of the 55 (NN,TRN) pairs, 44 are located within $500\,
\bp$ from each other, which is much smaller than the mean spacing of
$9.6\,\kbp$ between operons. Fig.~\ref{fig:Ftrn} also shows that for
the ensemble of \trn\ pairs that are \emph{not} nearest neighbours,
the cumulative distribution function is much closer to the average of
the ensemble of random networks. This establishes that the bias towards short
distances in the \trn\ is due to the tendency of network pairs to be
nearest neighbours.

\begin{table*}[t]
\begin{tabular}{ll|rdl|rdl}  
  & & \multicolumn{3}{c|}{all pairs} 
  &   \multicolumn{3}{c}{pairs ($s<500\,\bp$)} \\[0.5ex]
Set & Orientation & \Ecoli\ & \multicolumn{1}{c}{Random} & $P$-value
                  & \Ecoli\ & \multicolumn{1}{c}{Random} & $P$-value \\
\hline 
Regulation Network (TRN) &&  497 &&&  45 && \\[0.15cm]
Coregulated (CR)        && 4362 &&&  27 && \\[0.15cm]
Autoregulated (AR)      &&  318 &&&  23 && \\[0.15cm]
TRN \& CR (ie FFls)      &&   42 &&&   6 && \\[0.15cm]
AR \& CR                 &&   24 &&&   5 && \\[0.15cm]
Nearest Neighbours (NN) %
  & Diverging   &  103 & & &  45 &  4.8 & $10^{-28}$ \\
(I-VI)  & Converging  &  105 & & &  13 &  4.8 & $10^{-3}$  \\
  & Tandem      &  188 & & &  42 &  9.6 & $10^{-14}$ \\[0.15cm]
TRN \& NN %
  & Diverging  & 33 & 0.63 & $10^{-44}$ & 27 & 0.031 & $10^{-69}$ \\
(III-VI)  & Converging &  3 & 0.63 & $10^{-2}$  &  2 & 0.031 & $10^{-4}$  \\
  & Tandem     & 19 & 1.3  & $10^{-16}$ & 15 & 0.061 & $10^{-31}$ \\[0.15cm]
CR \& NN %
  & Diverging  & 22 &  5.5 & $10^{-7}$  & 15 & 0.27 & $10^{-21}$ \\
(II+III+VI)  & Converging & 12 &  5.5 & $10^{-2}$  &  2 & 0.27 & $10^{-2}$  \\
  & Tandem     & 26 & 11.  & $10^{-4}$  &  8 & 0.54 & $10^{-7}$  \\[0.15cm]
AR \& NN %
  & Diverging  & 21 & 0.40 & $10^{-28}$ & 18 & 0.020 & $10^{-47}$ \\
(V+VI)  & Converging &  1 & 0.40 & 0.3        &  1 & 0.020 & $10^{-2}$  \\
  & Tandem     &  5 & 0.81 & $10^{-3}$  &  4 & 0.040 & $10^{-7}$  \\[0.15cm]
TRN \& CR \& NN %
  & Diverging  &  6 & 0.054 & $10^{-11}$ &  4 & 0.0026 & $10^{-12}$ \\
(III+VI)  & Converging &  1 & 0.054 & $10^{-2}$  &  1 & 0.0026 & $10^{-3}$  \\
  & Tandem     &  1 & 0.11  & 0.1        &  1 & 0.0052 & $10^{-2}$  \\[0.15cm]
\end{tabular}
\caption[]{\label{tab:overview} Sizes of (sub)sets shown in
Fig.~\ref{fig:overview} for \Ecoli; the roman numerals between
brackets refer to the regions in Fig.~\ref{fig:overview}.  The
quantities in the `Random' column refer to averages in the ensemble of
random networks as defined in Method.  The $P$-values are
probabilities that a quantity of at least the size as observed in the
\Ecoli\ network can be found in a random network.  The TRN \& CR \& NN
(III+VI) is identical to the AR \& CR \& NN set (VI) (not shown) apart
from the additional presence of one converging operon pair.}
\end{table*}

\mysection{Neighboring operons on the DNA}
The results on the separation statistics for the \trn\ motivated us to
examine the nearest neighbour pairs in more detail. We did not only
include pairs that constitute links in the \trn\ (regions III-VI) in
Fig.~\ref{fig:overview}, but also pairs that are coregulated by a
common transcription factor (regions II,III and VI in
Fig.~\ref{fig:overview}).

\begin{figure}[t]
\includegraphics[width=8.5cm]{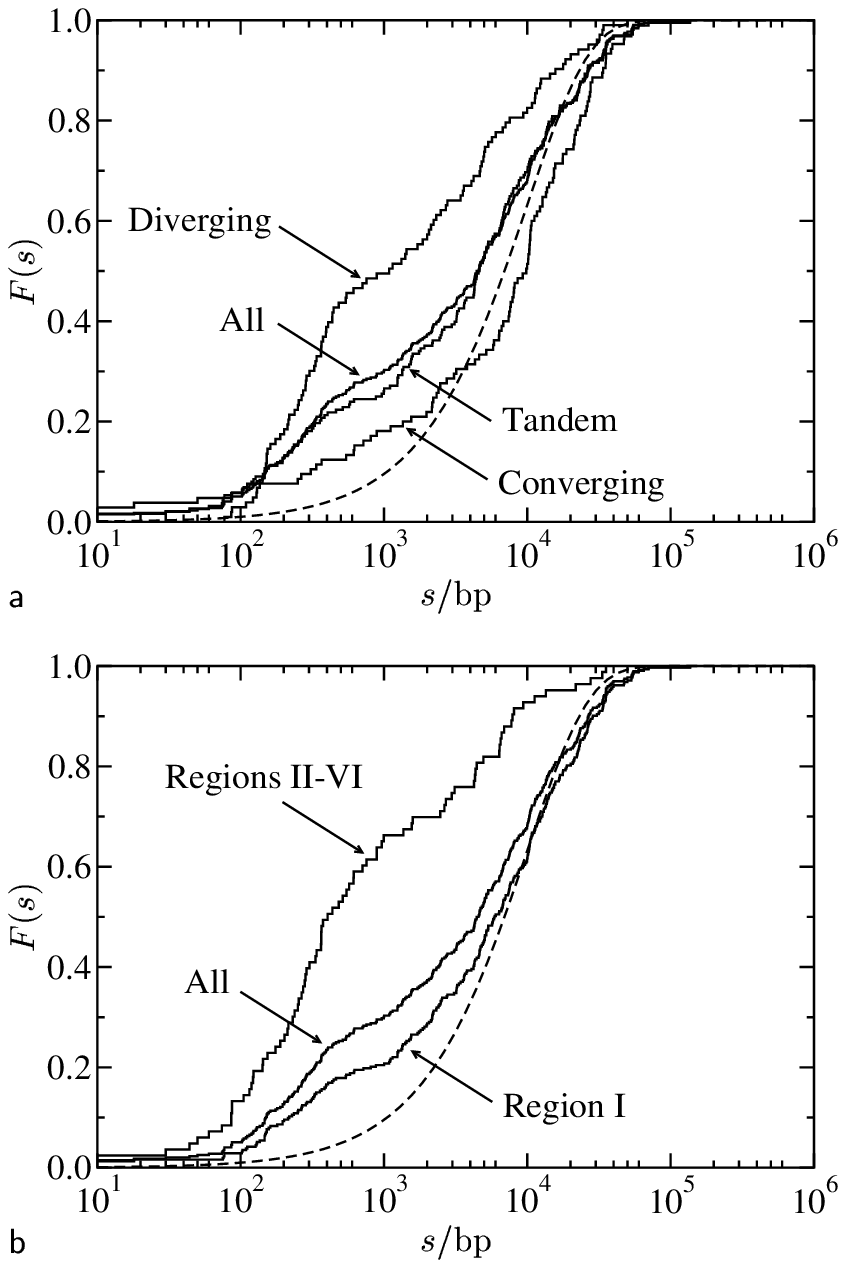}
\caption[]{The cumulative distribution functions $F(s)$ for nearest
neighbour pairs (NN pairs) in the \Ecoli\ network: (a) split by
relative orientation; and (b) split according to whether the neighbour
pair is coregulated or in the \trn\ (regions II--VI), or not (region
I).  The dashed line is given by a Poisson distribution,
$F(s)=1-\exp[-s/\Lrand]$, where $\Lrand=9.6\,\kbp$ is the mean spacing
between operons.\label{fig:neighbour}}
\end{figure}

Table~\ref{tab:overview} shows the sizes of the unions formed by the
overlap of, on the hand, the set of nearest neighbour (NN) pairs, and,
on the other hand, the sets of (autoregulatory) network pairs and
coregulated pairs, respectively. It is seen that these unions are significantly
larger than the corresponding unions in the ensemble of random
networks. Moreover, region I, corresponding to the ensemble of nearest
neighbours that are neither coregulated nor form a link in the \trn,
is {\em smaller} than the corresponding region in the ensemble of
random networks (data not shown). Clearly, both operons that regulate each
other and operons that are coregulated by a common transcription
factor, tend to be nearest neighbours on the DNA.

A spatial arrangement of operons in which adjacent operons are
transcribed in diverging directions, allows the upstream regulatory
regions to overlap. As we discuss in more detail below, this affords
additional regulatory control. We therefore addressed two questions:
1) Do operons that are each other's nearest neighbours tend to be
transcribed in diverging directions? 2) Do these nearest neighbours
tend to lie relatively close to each other?

Table~\ref{tab:overview} shows the statistics for pairs of operons
broken down according to the relative direction of
transcription. There are three classes: `tandem' (both operons are
transcribed in a common direction), `converging' and `diverging'. If
the three classes of orientation were populated in a random manner,
one would expect the ratio converging : diverging : tandem = 1 : 1 :
2. Our analysis reveals that in region I and in regions VII - XI, the
three classes are indeed populated in a nearly random manner (data not
shown). In contrast, the orientation statistics for adjacent operons
that are either coregulated or form a (autoregulatory) link in the
\trn, are markedly different. These pairs tend to be transcribed in
diverging directions. Moreover, this effect is most pronounced if the
operons lie very close to each other on the
DNA. Table~\ref{tab:overview} shows that coregulated and
(autoregulatory) network pairs that lie less than $500 \, \bp$ apart
from each other, are predominantly transcribed in diverging
directions.

To answer the second question, we have calculated the
separation-distribution function $F(s)$ for nearest neighbour
pairs. If operons were distributed at random on the genome, one would
expect the separation statistics to follow a Poisson distribution;
this corresponds to a model in statistical physics known as a Tonks
gas~\cite{phys1d}. Fig.~\ref{fig:neighbour}, however, shows that the
\Ecoli\ network exhibits strong deviations from Poisson statistics. In
particular, it shows that a large number of links are distinctively
short. This is most striking for operons that are transcribed in
diverging directions, although it is also noticeable for operons that
are transcribed in a common direction. About 45\% of diverging pairs
and 20\% of tandem pairs are closer than $500\,\bp$, and these
fractions are much higher than would be expected for a random network
(see also Table~\ref{tab:overview}). It appears that the \trn\ of
\Ecoli\ has a large fraction of adjoining operons. Importantly,
Fig.~\ref{fig:neighbour}b indicates that most of these adjoining
operons are, indeed, operons that either regulate each other or are controlled
by a common transcription factor.

As mentioned above, an arrangement in which neighboring operons are
transcribed in diverging directions, allows the operator regions to
interfere with each other.  The occurrence of diverging neighbour
pairs with operator interference can be assessed by a careful
examination of the EcoCyc database \cite{note1,ecocyc} (for details
see supplementary information). Of the 45 diverging neighbour pairs
with $s<500\,\bp$ (see Table~\ref{tab:overview}), 20 operon pairs have operator
interference (there are, in fact, 3 examples of operator interference
with $s>500\,\bp$); 10 do not; and for the remaining 15, there is
insufficient information on the promoter/operator regions to decide.
We conclude that the presence of operator interference provides a
major part of the explanation for the strong bias towards small
separations for diverging pairs.

\mysection{Discussion}
The principal findings of our statistical analysis are: 1) pairs of
operons that regulate each other and pairs of operons that are
coregulated tend to be nearest neighbours; 2) these nearest neighbours
tend to be transcribed in diverging directions; 3) the nearest
neighbours' separation statistics is strongly biased towards short
distances.  What could be the origin of this behavior?  Two distinct
scenarios can give rise to the observed separation statistics.  In the
first, the bias is a result of the mechanisms by which new operons and
new links between operons emerge and disappear in the course of
evolution.  In the second, it is a consequence of a functional benefit
for having short separations between certain pairs of operons; in this
scenario, there is a selection pressure towards shorter distances.  It
seems hard, if not impossible, to disentangle both mechanisms,
although, in principle, there is a clear difference between the two.
In the former scenario, newly emerged operons will drift apart and, as
a function of time, the distance between them will increase.  In
contrast, in the latter scenario, a selection pressure will drive and
keep the operons together.

Still, the question remains why it could be beneficial to have pairs of
operons together.  It is believed that in prokaryotes transcription
and translation take place simultaneously (see Fig. 1.2 in
Wagner~\cite{trpbook}).  It seems advantageous to have transcription
factors expressed close to the locus at which they are supposed to
act.  Furthermore, two operons can be topologically coupled via the
interplay between transcription and supercoiling~\cite{trpbook}.  This
can lead to additional regulatory control, but only if the distance
between the operons is covered by the twin-supercoiled domains.

\begin{figure}[t]
\includegraphics[width=8.5cm]{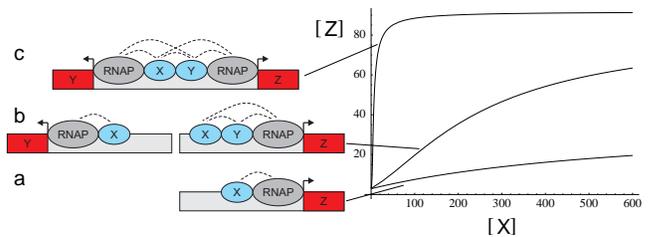}
\caption[]{Response of FFLs as a function of the spatial arrangement
of the regulatory elements.  The transcription factors X and Y
coherently regulate the expression of operon Z; a dashed line
indicates a weak cooperative interaction of $\approx 3 k_BT$, which
corresponds to a cooperativity factor $\omega \approx 20$; RNAP
denotes the enzyme RNA polymerase.  On the left
the different structures: {\bf a} the expression of gene Z is
activated by the transcription factor X only; {\bf b} a `classical'
FFL; {\bf c} a FFL in which the operator regions overlap. On the
right, the concentration in nM of the expressed protein Z as a
function of that of the transcription factor X; the inducer for
transcription factor Y is assumed to be present at saturating
concentrations~\cite{Mangan03}. It is seen that a FFL can act as an
amplifier and that overlapping operons can significantly enhance the
performance of the amplifier.\label{fig:FFW}}
\end{figure}

Perhaps the most interesting case concerns the prevalence for short
distances amongst neighboring operons that are transcribed in
diverging directions.  This spatial arrangement of operons offers the
possibility that the operator regions interfere with each other.  This
can provide an extra layer of regulatory control.  Just as the
existence of operons provides for correlated \emph{gene} expression,
interference between the regulatory regions for a pair of diverging
operons affords additional opportunities for correlated or
\emph{anticorrelated} expression of \emph{operons}.

One of the simplest regulatory constructs is a genetic switch
consisting of two operons that mutually repress each other, and
elsewhere we will publish a detailed analysis of the consequences of
correlated and anticorrelated operon expression for the stability of
such toggle switches~\cite{note-inprep}.  Toggle switches are not a
statistically significant motif in the \trn\ of \Ecoli\ though, and to
demonstrate the effect we turn to an example based around a
feed-forward loop (FFL). Shen-Orr \etal~\cite{SOMMA} demonstrated that
FFLs are important computational elements of the \Ecoli\ network and
it is believed that they can perform a variety of computational tasks
in a regulatory circuit. It is believed that they can filter transient
signals~\cite{SOMMA,Mangan03}, act as sign-sensitive accelerators or
sign-sensitive delays~\cite{Mangan03}, or act as an amplifier, in
which the activity of the gene at the top of the loop is amplified at
the ultimate target gene~\cite{Mangan03,Lee02}.

Our statistical analysis has revealed that in FFLs the downstream
operons tend to overlap (see Table~\ref{tab:overview}). In order to
investigate the role of overlapping operons in FFLs, we have performed
a mean-field analysis of FFLs in which transcription factors X and Y
coherently activate the expression of operon Z~\cite{Mangan03} (see
Fig.~\ref{fig:FFW} and supplementary material for details).
Fig.~\ref{fig:FFW} shows that overlapping operons can strongly enhance
the sharpness of the response. In contrast to the general scheme
(Fig.~\ref{fig:FFW}b), the gene regualtory proteins X and Y can {\em
simultaneously} activate gene Y and gene Z in the overlapping operon
scenario (Fig.~\ref{fig:FFW}c).  This allows for extra cooperativity,
which in turn leads to a sharper response.
It would seem that the capacity to generate a strong reponse can
confer a competitive advantage to the organism in a number of cases,
such as in the repression of sugar-uptake systems in response to
glucose. Hence, our results suggest that regulatory control can
provide a selection pressure that drives operons together in the
course of evolution.

\mysection{Acknowledgements}
We thank Marileen Dogterom, Daan Frenkel, Sander Tans and Conrad
Woldringh for useful discussions and a careful reading of the
manuscript. The work is supported by the Amsterdam Centre for
Computational Science (ACCS). This work is part of the research
program of the `Stichting voor Fundamenteel Onderzoek der Materie
(FOM)', which is financially supported by the `Nederlandse organisatie
voor Wetenschappelijk Onderzoek (NWO)'.




\clearpage
\onecolumngrid

\mysection{Supplementary material -- overlapping operons}
\begin{center}
\small
\begin{ruledtabular}
\begin{tabular}{lclcrrr}
{\bf operon 1} & {\bf start (dirn) end} &
{\bf operon 2} & {\bf start (dirn) end} &
{\bf sepn} & {\bf (sub)set} & {\bf interference}\\
\hline
mhpABCDFE    &  367835 ($+$) 374105  & 
mhpR         &  366811 ($-$) 367758  &   77 &      TRN & unknown \\
soxS         & 4274639 ($-$) 4274962 &
soxR         & 4275048 ($+$) 4275512 &   86 &       AR & yes \\
bioA         &  807191 ($-$) 808480  & 
bioBFCD      &  808567 ($+$) 812170  &   87 &       CR & yes \\
cynTSX       &  358023 ($+$) 360370  &
cynR         &  357015 ($-$) 357914  &  109 &       AR & yes \\
lysA         & 2975659 ($-$) 2976921 & 
lysR         & 2977043 ($+$) 2977978 &  122 &       AR & yes \\
\hline
betT         &  328687 ($+$) 330720  &
betIBA       &  324801 ($-$) 328558  &  129 &       AR & yes \\
torCAD       & 1057307 ($+$) 1061621 &
torR         & 1056485 ($-$) 1057177 &  130 &       AR & yes \\
hcaA1A2CBD-y & 2667052 ($+$) 2671788 & 
hcaR         & 2666026 ($-$) 2666916 &  136 &       AR & unknown \\
acrAB        &  480478 ($-$) 484843  & 
acrR         &  484985 ($+$) 485632  &  142 &      TRN & unknown \\
gyrA         & 2334813 ($-$) 2337440 & 
ubiG         & 2337587 ($+$) 2338309 &  147 &     none & no \\
\hline
cpxP         & 4103398 ($+$) 4103900 &
cpxAR        & 4101183 ($-$) 4103251 &  147 &       AR & unknown \\
ilvC         & 3955591 ($+$) 3957066 & 
ilvY         & 3954548 ($-$) 3955441 &  150 &       AR & yes \\
pspABCDE     & 1366103 ($+$) 1368027 & 
pspF         & 1364959 ($-$) 1365951 &  152 &      TRN & yes \\
asnA         & 3924783 ($+$) 3925775 & 
asnC         & 3924173 ($-$) 3924631 &  152 &       AR & no \\
argCBH       & 4152580 ($+$) 4155802 & 
argE         & 4151275 ($-$) 4152426 &  154 &       CR & yes \\
\hline
iclMR        & 4220383 ($-$) 4221246 & 
metH         & 4221407 ($+$) 4225090 &  161 &     none & unknown \\
malXY        & 1697379 ($+$) 1700153 & 
malI         & 1696176 ($-$) 1697204 &  175 & AR \& CR & yes \\
rtcAB        & 3554044 ($-$) 3555711 & 
rtcR         & 3555900 ($+$) 3557498 &  189 &      TRN & unknown \\
yiaKLMNOPQRS & 3740362 ($+$) 3744710 & 
yiaJ         & 3739313 ($-$) 3740161 &  201 &      TRN & yes \\
hycABCDEFGH  & 2841059 ($-$) 2848458 & 
hypA         & 2848670 ($+$) 2849020 &  212 &       CR & yes \\
\hline
fliE         & 2010722 ($-$) 2011036 & 
fliFGHIJK    & 2011251 ($+$) 2017535 &  215 &       CR & unknown \\
dsdXA        & 2475867 ($+$) 2478550 & 
dsdC         & 2474714 ($-$) 2475649 &  218 &       AR & no \\
zraP         & 4198841 ($-$) 4199266 & 
hydHG        & 4199504 ($+$) 4202223 &  238 &      TRN & unknown \\
glcDEFGB     & 3119650 ($-$) 3126036 & 
glcC         & 3126287 ($+$) 3127051 &  251 &      TRN & yes \\
ssb          & 4271704 ($+$) 4272240 & 
uvrA         & 4268628 ($-$) 4271450 &  254 &       CR & yes \\
\hline
fliC         & 2000133 ($-$) 2001629 & 
fliDST       & 2001895 ($+$) 2004101 &  266 &       CR & unknown \\
glpACB       & 2350667 ($+$) 2354731 & 
glpTQ        & 2347955 ($-$) 2350394 &  273 &       CR & yes \\
melAB        & 4339489 ($+$) 4342368 & 
melR         & 4338298 ($-$) 4339206 &  283 & AR \& CR & yes \\
rhaT         & 4097072 ($-$) 4098106 & 
sodA         & 4098391 ($+$) 4099011 &  285 &     none & no \\
rhaBAD       & 4091029 ($-$) 4095029 & 
rhaSR        & 4095317 ($+$) 4097075 &  288 &       AR & no \\
\hline
yhfA         & 3483051 ($-$) 3483455 & 
crp          & 3483757 ($+$) 3484389 &  302 &       AR & yes \\
rpiB         & 4310929 ($+$) 4311378 & 
rpiR-alsBACE & 4309680 ($-$) 4310603 &  326 &       AR & unknown \\
nagE         &  703167 ($+$) 705113  & 
nagBACD      &  698797 ($-$) 702834  &  333 & AR \& CR & no \\
araBAD       &   65855 ($-$) 70048   & 
araC         &   70387 ($+$) 71265   &  339 & AR \& CR & yes \\
exuT         & 3242744 ($+$) 3244162 & 
uxaCA        & 3239467 ($-$) 3242381 &  363 &       CR & unknown \\
\hline
malEFG       & 4240205 ($-$) 4243998 & 
malK-lamB-ma & 4244363 ($+$) 4248053 &  365 &       CR & yes \\
xylAB        & 3725546 ($-$) 3728394 & 
xylFGHR      & 3728760 ($+$) 3733786 &  366 &       AR & no \\
entCEBA      &  624108 ($+$) 628520  & 
fepB         &  622777 ($-$) 623733  &  375 &       CR & no \\
acs          & 4282992 ($-$) 4284950 & 
nrfABCDEFG   & 4285343 ($+$) 4291718 &  393 &     none & unknown \\
purL         & 2689676 ($-$) 2693563 & 
yfhD         & 2693959 ($+$) 2695377 &  396 &     none & unknown \\
\hline
nadB         & 2708440 ($+$) 2710062 & 
rpoE-rseABC  & 2705342 ($-$) 2708032 &  408 &     none & no \\
argR         & 3382338 ($+$) 3382808 & 
mdh          & 3380965 ($-$) 3381903 &  435 &     none & no \\
caiTABCDE    &   34781 ($-$) 41931   & 
fixABCX      &   42367 ($+$) 45750   &  436 &       CR & yes \\
modABC       &  794312 ($+$) 796835  & 
modE         &  793079 ($-$) 793867  &  445 &      TRN & unknown \\
leuLABCD     &   78848 ($-$) 83708   & 
leuO         &   84191 ($+$) 85312   &  483 &      TRN & unknown \\
\hline
aroP         &  120178 ($-$) 121551  & 
pdhR-aceEF-l &  122092 ($+$) 129336  &  541 &     none & no \\
fucAO        & 2929887 ($-$) 2931710 & 
fucPIKUR     & 2932257 ($+$) 2938121 &  547 & AR \& CR & no \\
malPQ        & 3545619 ($-$) 3550106 & 
malT         & 3550718 ($+$) 3553423 &  612 &      TRN & no \\
gltA         &  752408 ($-$) 753691  & 
sdhCDAB-b072 &  754400 ($+$) 764272  &  709 &       CR & yes \\
csgBA        & 1103174 ($+$) 1104125 & 
csgDEFG      & 1100074 ($-$) 1102419 &  755 & AR \& CR & no \\
\hline
flgBCDEFGHIJ & 1130241 ($+$) 1139244 & 
flgMN        & 1128637 ($-$) 1129351 &  890 &       CR & unknown \\
glpD         & 3559646 ($+$) 3561151 & 
glpR         & 3557480 ($-$) 3558238 & 1408 &      TRN & yes \\
narK         & 1277180 ($+$) 1278571 & 
narL         & 1274402 ($-$) 1275052 & 2128 &      TRN & yes \\
\end{tabular}
\end{ruledtabular}
\end{center}

\noindent Included are all diverging neighbour pairs with less than
$1\,\kbp$ separation between coding regions, or with detected
interference between operator regions if separation is greater than
$1\,\kbp$.  Operon names have been truncated to 12 characters to save
space.  All the AR \& CR pairs are also examples of `downstream'
operons in feed-forward loops.  For TRN and AR pairs, operon 1 is the
controlled operon and operon 2 is the controlling operon.

\clearpage

\mysection{Supplementary material -- feed-forward loops}

\noindent We follow the approach of Shea and Ackers~\cite{Shea85} and
Buchler et al.~\cite{Buchler03} in modelling gene expression. The
approach  relies on the idea of ``regulated
recruitment''~\cite{Ptashne97,gensigbook}: gene regulatory proteins
control gene expression by modulating the probability $P$ that the
enzyme RNA polymerase is bound to the DNA; if the RNA polymerase is
bound, then it is assumed that gene expression occurs at a fixed rate
$\beta$. The macroscopic rate equation for the synthesis and
degradation of a protein ${\rm Z}$ is thus given by:
\begin{equation}
\frac{{\rm d}[{\rm Z}]}{{\rm dt}} = \beta P_{\rm Z}  - \mu [{\rm Z}].
\end{equation}
Here $[{\rm Z}]$ is the concentration of protein ${\rm Z}$, $P_{\rm Z}$ is the
probability that the RNA polymerase is bound to the promoter for gene
${\rm Z}$ and $\mu$ is the degradation rate of the protein. The
probability $P_{\rm Z}$ that a RNA polymerase is bound to the promoter
of gene ${\rm Z}$
is given by
\begin{eqnarray}
P_{\rm Z} &=&  \frac{Z_{{\rm Z}, \rm on}}{Z_{{\rm Z},\rm
off} + Z_{{\rm Z}, \rm on}}.
\end{eqnarray} 
Here $Z_{{\rm Z}, \rm on}$ and $Z_{{\rm Z}, \rm off}$ are the partition functions
for the system with the RNA polymerase bound and not bound to the
promoter of gene ${\rm Z}$, respectively. Following Buchler et
al.~\cite{Buchler03}, we characterize the interaction between a pair
of proteins  -- a protein being either a RNA polymerase or a
transcription factor -- by a
cooperativity factor $\omega$. A weak glue-like interaction of $3 k_B
T \approx 2 \mbox{kcal/mol}$ is assumed~\cite{Buchler03,Ptashne97,gensigbook},
which corresponds to a cooperativity factor $\omega \approx 20$. 
Furhermore, we assume that if
gene expression is controlled by two gene regulatory proteins, the RNA
polymerase can contact both proteins simultaneously. We have also
considered the independent interaction model, in which the RNA
polymerase can only interact with one gene regulatory protein at the
time~\cite{Buchler03}. We obtained similar conclusions for the two models.

We now consider the structures shown in Fig.~\ref{fig:FFWsup} of the
supplementary material.

\noindent{\bf Structure a:}
\begin{eqnarray}
Z_{{\rm Z}, {\rm off}} &=& 1+[{\rm X}]/K_{\rm X} \nonumber \\
Z_{{\rm Z}, {\rm on}} &=& [{\rm RNAP}]/K_{\rm RNAP} \left (1 + \omega [{\rm X}] / K_{\rm X}\right)\nonumber
\end{eqnarray}
\noindent{\bf Structure b:}
\begin{eqnarray}
Z_{{\rm Y},{\rm off}} &=& 1+[{\rm X}]/K_{\rm X}\nonumber \\
{\rm Z}_{{\rm Y}, {\rm on}} &=& [{\rm RNAP}]/K_{\rm RNAP} \left(1 + 
			                    \omega [{\rm X}] / K_{\rm
X}\right)\nonumber \\
Z_{{\rm Z}, {\rm off}} &=& 1+[{\rm X}]/K_{\rm X} + [{\rm Y}]/K_{\rm Y} + \omega [{\rm X}][{\rm Y}]/(K_{\rm X} K_{\rm Y})\nonumber \\
Z_{{\rm Z}, {\rm on}} &=& [{\rm RNAP}]/K_{\rm RNAP} \left(1 + \omega [{\rm X}]/K_{\rm X} +
\omega [{\rm Y}] / K_{\rm Y} + \omega^3 [{\rm X}][{\rm Y}]/(K_{\rm X} K_{\rm Y})\right).\nonumber
\end{eqnarray}
\noindent{\bf Structure c:}
\begin{eqnarray}
Z_{{\rm Y}, {\rm off}} &=& 1+[{\rm X}]/K_{\rm X} + [{\rm Y}]/K_{\rm Y} + \omega [{\rm X}][{\rm Y}]/(K_{\rm X} K_{\rm Y})\nonumber \\
{\rm Z}_{{\rm Y}, {\rm on}} &=& [{\rm RNAP}]/K_{\rm RNAP} \left(1 + \omega [{\rm X}]/K_{\rm X} +
\omega [{\rm Y}] / K_{\rm Y} + \omega^3 [{\rm X}][{\rm Y}]/(K_{\rm X} K_{\rm Y})\right)\nonumber \\
Z_{{\rm Z}, {\rm off}} &=& Z_{{\rm Y}, {\rm off}}\nonumber \\
Z_{Z, {\rm on}} &=& Z_{{\rm Y}, {\rm on}}\nonumber
\end{eqnarray}
\noindent It is seen that the expression of ${\rm Y}$ depends on the
concentration of ${\rm Y}$. This means that in order to obtain the
concentration of ${\rm Y}$, we have to solve a quadratic equation in
$[{\rm Y}]$. This structure is included to show the effect of the
autoregulatory loop on {\rm Y} -- the transcription factor {\rm Y}
interacts with the RNA polymerase bound to the promoter for gene {\rm
Y}. \\[0.25cm]
\noindent{\bf Structure d:}
\begin{eqnarray}
Z_{{\rm Y}, {\rm off}} & = & 1 + [{\rm X}]/K_{\rm X} + [{\rm Y}]/K_{\rm Y} + \omega [{\rm X}][{\rm Y}]/(K_{\rm X} K_{\rm Y})
+ [{\rm RNAP}]/K_{\rm RNAP} + \omega [{\rm X}][{\rm RNAP}] / (K_{\rm X} K_{\rm
RNAP}) \nonumber \\
 && + \omega [{\rm Y}] [{\rm RNAP}]/ (K_{\rm Y} K_{\rm RNAP}) +  \omega^3 [{\rm X}][{\rm Y}][{\rm RNAP}]/(K_{\rm X} K_{\rm Y} K_{\rm RNAP})\nonumber \\ 
Z_{{\rm Y},{\rm on}} &=& [{\rm RNAP}] / K_{\rm RNAP} \left(1+\omega [{\rm X}]/K_{\rm X} + \omega [{\rm Y}]/K_{\rm Y} + [{\rm
RNAP}]/K_{\rm RNAP} +\right. \nonumber \\
&& \omega^3 [{\rm X}][{\rm Y}]/(K_{\rm X} K_{\rm Y}) + \omega^2 [{\rm X}][{\rm
RNAP}] / (K_{\rm X} K_{\rm RNAP}) + \omega^2 [{\rm Y}][{\rm RNAP}] / (K_{\rm Y} K_{\rm
RNAP})\nonumber \\
&&\left. + \omega^5 [{\rm X}][{\rm Y}][{\rm RNAP}]/(K_{\rm X} K_{\rm Y} K_{\rm RNAP})\right)\nonumber \\
Z_{{\rm Z}, {\rm off}} &= & Z_{{\rm Y}, {\rm off}}\nonumber \\
Z_{{\rm Z}, {\rm on}} &=& Z_{{\rm Y}, {\rm on}}\nonumber
\end{eqnarray}
\noindent Note that this structure corresponds to structure {\bf c} of
Fig.~\ref{fig:FFW} in the main text.

We have taken $K_{\rm X} = K_{\rm Y} = K_{\rm RNAP} = 1000 \, {\rm
nM}$, where, for {\em E. coli}, $1{\rm nM}$ corresponds to roughly one
molecule per cell. Note that structure {\bf c} of Fig.~\ref{fig:FFW}
of the main text corresponds to structure {\bf d} of
Fig.~\ref{fig:FFWsup} of the supplementary material.

\begin{figure}[t]
\includegraphics[width=16cm]{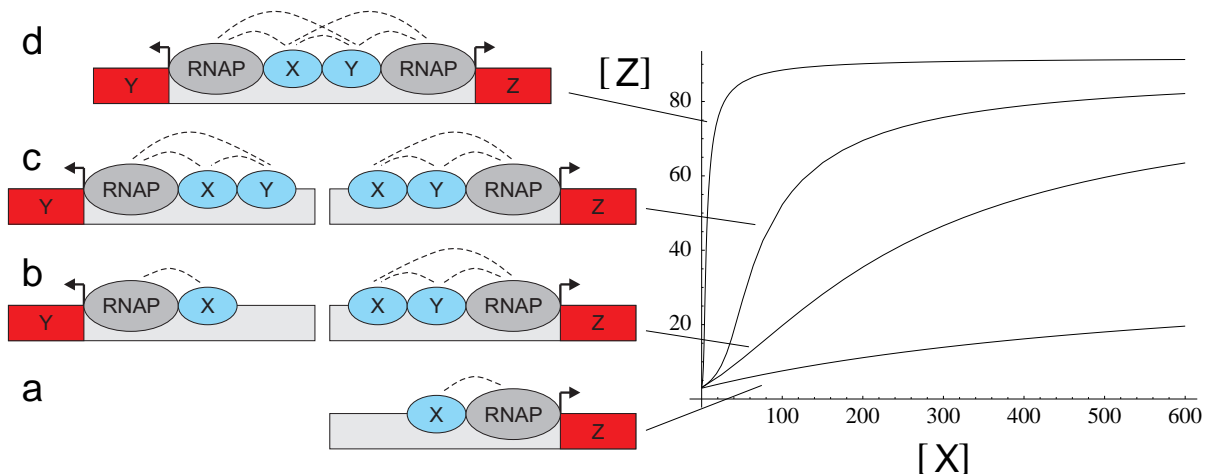}
\caption{\label{fig:FFWsup} Response of FFLs as a function of the
spatial arrangement of the regulatory elements. A dashed line
indicates a weak cooperative interaction of $\approx 3 k_BT$, which
corresponds to a cooperativity factor $\omega \approx 20$. On the
left, the different strucutres: {\bf a} the expression of gene Z is
activated by the transcription factor X only; {\bf b} a ``classical''
FFL; {\bf c} a FFL with an autoregulatory loop on Y; {\bf d} a FFL in
which the operator regions overlap. On the right, the concentration of
the expressed protein Z in nM as a function of that of the
transcription factor X. Note that structure {\bf d} corresponds to
structure {\bf c} of Fig.~\ref{fig:FFW} of the main text. It is seen
that a FFL can act as an amplifier and that overlapping operons can
significantly enhance the performance of the amplifier.}
\end{figure}

We have also performed an extra set of calculations, in which there is
no direct interaction between the transcription factor Y and the RNA
polymerase bound to the promoter for gene Y. For all sets of
calculations, we found that the structure in which the operator
regions overlap (i.e. structure {\bf c} of Fig.~\ref{fig:FFW} of the
main text and structure {\bf d} of Fig.~\ref{fig:FFWsup} of the
supplementary material) gives the sharpest response.

\end{document}